\documentclass[10pt,letterpaper]{article}

\usepackage{fullpage}

\usepackage[utf8]{inputenc}

\usepackage{amsmath,amssymb}

\usepackage{cite}

\usepackage{rotating}

\usepackage{multirow}

\usepackage[aboveskip=1pt,labelfont=bf,labelsep=period,singlelinecheck=off]{caption}

\makeatletter
\renewcommand{\@biblabel}[1]{\quad#1.}
\makeatother

\date{}

\usepackage{graphicx}

\begin{document}
\vspace*{0.35in}

\begin{center}
{\Large
\textbf{Theory and associated phenomenology for\\intrinsic mortality arising from natural selection}
}
\newline
\\
\renewcommand{\thefootnote}{\fnsymbol{footnote}}
Justin Werfel\textsuperscript{1,2,3,}\footnotemark,
Donald E.\ Ingber\textsuperscript{1,3,4},
Yaneer Bar-Yam\textsuperscript{2}
\\
\bigskip
\bf{1} Wyss Institute for Biologically Inspired Engineering,\\Harvard University, Cambridge, MA, USA
\\
\bf{2} New England Complex Systems Institute, Cambridge, MA, USA
\\
\bf{3} Harvard Medical School and Children's Hospital, Boston, MA, USA
\\
\bf{4} School of Engineering and Applied Sciences, Harvard University, Cambridge, MA, USA
\\
\bigskip

\footnotetext{justin.werfel@wyss.harvard.edu}

\end{center}

\section*{Abstract}
Standard evolutionary theories of aging and mortality, implicitly
based on assumptions of spatial averaging, hold that natural selection
cannot favor shorter lifespan without direct compensating benefit to
individual reproductive success.  Here we show that both theory and
phenomenology are consistent with programmed death.  Spatial
evolutionary models show that self-limited lifespan robustly results
in long-term benefit to a lineage; longer-lived variants may have a
reproductive advantage for many generations, but shorter lifespan
ultimately confers long-term reproductive advantage through
environmental feedback acting on much longer time scales.  Numerous
model variations produce the same qualitative result, demonstrating
insensitivity to detailed assumptions; the key conditions under which
self-limited lifespan is favored are spatial extent and locally
exhaustible resources.  Numerous empirical observations can
parsimoniously be explained in terms of long-term selective advantage
for intrinsic mortality.  Classically anomalous empirical data on
natural lifespans and intrinsic mortality, including observations of
longer lifespan associated with increased predation, and evidence of
programmed death in both unicellular and multicellular organisms, are
consistent with specific model predictions.  The generic nature of the
spatial model conditions under which intrinsic mortality is favored
suggests a firm theoretical basis for the idea that evolution can
quite generally select for shorter lifespan directly.

\section*{Author Summary}
Classic evolutionary theories hold that selection cannot directly favor reduced lifespan. Intuitively, selection should act against a gene contributing to the death of its owner. Mortality is instead explained as an indirect effect of traits with net benefit to the individual. However, these theories are implicitly based on averaging over populations. Spatial systems, in which heterogeneity can arise, have qualitatively different behavior. We show that in spatial population models, self-limited lifespan is favored even without direct compensating benefit, and that this phenomenon can help explain many observations in nature. In the models, longer lifespan is favored on short time scales, but environmental feedback based on heterogeneous resource use results in shorter-lived strains having a long-term advantage. This result can help to explain observations that have been surprising or difficult to explain with classic theories. These results add a key mechanism to our understanding of the evolution of lifespan.

\section*{Introduction}

Lifespan extension is a topic of longstanding broad interest, with a
long history of varied treatments advocated for those interested in
living longer.  Lifespans vary widely among species
\cite{Weismann1891,Finch90,Ger94,Love02} and can be experimentally
altered by genetic modification
\cite{KenyonEtAl93,LakowskiHekimi96,FinchTanzi97,GuarenteRuvkunAmasino98,Walker00,GuarenteKenyon00,TatarEtAl03,BudovskayaEtAl08,Kenyon10}
and selective breeding \cite{ZwaanEtAl95,RoseEtAl05}.  Standard
evolutionary theory explains senescence as the result of two core
effects, both based on the observation that the strength of selection
decreases as age increases, since sources of extrinsic mortality
reduce the older population size on which selection can act. The first
effect, mutation accumulation \cite{Medawar52,Charlesworth00},
describes traits with late-acting negative effects being only weakly
selected against and hence not eliminated. The second, antagonistic
pleiotropy \cite{Williams57} and its disposable soma formulation
\cite{Kirkwood77}, observes that traits conveying both early-life
benefits and later-life penalties can be actively selected for.  These
factors are generally considered to provide the complete basis for
senescence and intrinsic mortality, while older ideas of evolution of
direct lifespan control \cite{Weismann1891} are considered untenable;
there is a broadly accepted understanding that selection does not and
\emph{cannot} act directly in favor of shortening or otherwise
limiting lifespan
\cite{KirkwoodAustad00,OlshanskyEtAl02,HekimiGuarente03}. This
understanding accords with intuition: a gene that contributes to or
hastens the death of its owner should be eliminated.

However, a variety of observed phenomena have been and/or remain
difficult to reconcile with accepted frameworks
\cite{Kirkwood05}. These include large single-gene effects on lifespan
\cite{KenyonEtAl93} and other central regulators of aging
\cite{LakowskiHekimi96,TatarEtAl03,BudovskayaEtAl08}, semelparity
(one-time reproduction apparently triggering death, as in salmon)
\cite{Finch90,Ger94,MorbeyEtAl05}, anomalous patterns of longevity
\cite{Reznick04} and senescence \cite{Ger94,BaudischVaupel12}
including apparently nonaging species
\cite{Finch90,Ger94,Finch98,Dodd01} and even negative senescence
\cite{VaupelEtAl04,BaudischVaupel12}, aging in unicellular organisms
\cite{Ameisen96,Lewis00,DuszenkoEtAl06,KolodkinGalEtAl07}, and
significant plasticity of aging and lifespan
\cite{ZwaanEtAl95,RoseEtAl05}.  At the same time, more recent work in
spatial modeling of evolutionary systems has given new insight into
altruistic behaviors of individuals within populations \cite{Nowak06},
with relevance to phenomena including reproductive restraint
\cite{RauchEtAlPRL02,RauchEtAlJTB03}, social communication
\cite{pnas04}, and sexual reproduction
\cite{KeelingRand95,SalatheEtAl06}. Such models may thus help cast new
light on empirical observations related to aging.

\begin{figure}[tbp]
\begin{center}
\includegraphics[width=.8\columnwidth]{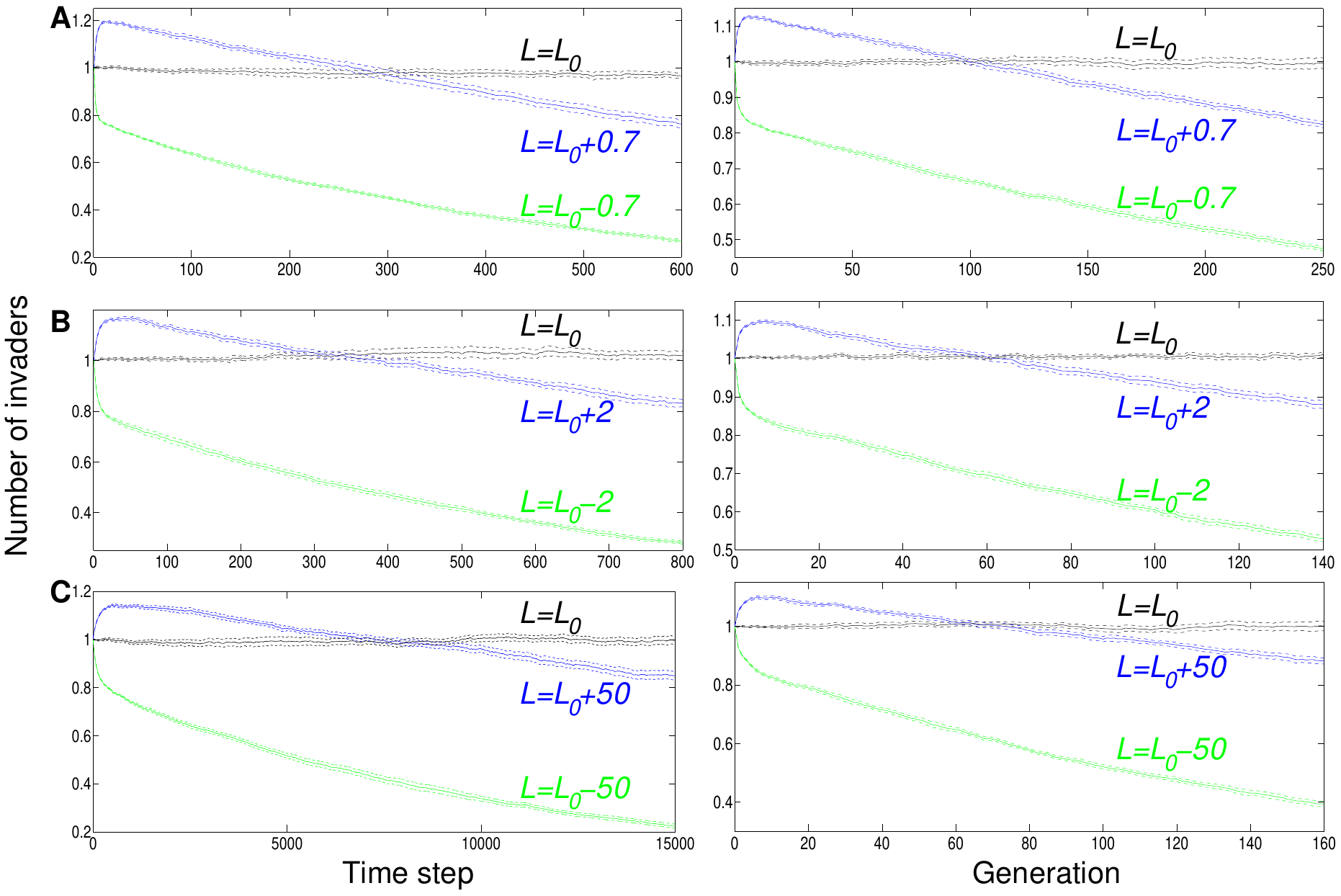}
\caption{{\bf Longer lifespan can give short-term advantage and
    long-term disadvantage in spatial models.} Plots show relative
  reproductive success (number of descendants) of individual mutants
  with lifespan $L$, introduced into a steady-state population with
  equilibrium lifespan $L=L_0$ (Fig.~\ref{fig:results}C), as a
  function of time (left) and of generation number (right) in
  simulations of a consumer-resource model (see text). Parameter
  values are (A) $g=0.13, v=0.2, L_0=2.7$, (B) $g=0.17, v=0.1,
  L_0=7.6$, (C) $g=0.2, v=0.005, L_0=170$, with resource growth rate
  $g$, consumption rate $v$, reproduction cost $c=0$, and no mutation
  ($\mu_p=\mu_q=0$) after the initial introduction. Error bars show
  the standard error of the mean among ten independent runs, each
  recording the mean for a set of 40,000 invasions.}
\label{timedep}
\end{center}
\end{figure}

In this paper we present and extend recent theoretical analysis
\cite{prl15} showing an adaptive mechanism for explicit control of
lifespan using spatial evolutionary models, and discuss the
application of the results to empirical observations.  In the models,
heterogeneity of limiting resources and self-organizing population
structures lead to an adaptive self-limitation of lifespan, without
direct compensating benefit to the individual.  The mechanism is
different from the standard evolutionary genetic theories founded on
mutation accumulation and antagonistic pleiotropy. In contrast to
these classic mechanisms, which are based on selection directly on the
reproductive success of individual organisms, this mechanism is based
in spatial heterogeneity and it predicts the evolution of traits that
may become advantageous only for descendants after many generations.
To make this counterintuitive result clear, we show explicitly
(Fig.~\ref{timedep}) that longer-lived mutants in prototypical
simulations have a temporary advantage that may last for hundreds of
generations, but eventually they are out-competed by the shorter-lived
types.  Thus, there is no compensating reproductive benefit to the
individual of the shortened lifespan, as there would have to be in a
mean-field model.  Our results show that the benefit that makes
shorter life favorable need not vest through reproductive success of
the individual, but rather through far descendants; moreover, the
mechanism is generic in spatial environments.

Why should spatial models be different from non-spatial ones?  The key lies in the short-term advantage but long-term disadvantage incurred by overly exploitative variants \cite{pnas04,RauchEtAlPRL02}.  ``Selfish'' variants that outcompete their neighbors, at the expense of depleting their local environment, have an advantage on short spatial and temporal scales. However, on longer scales, their descendants are left in impoverished environments and can be outcompeted by others in richer areas.  A longer-range view of ``time-dependent fitness'' \cite{RauchEtAlPRL02} recognizes these effects, and predicts the broad success of altruism and individual restraint in spatial models. Non-spatial models, including age-structured population models, examine the evolution of mutations affecting lifespan in large, homogenous populations. They have not studied the effects of spatial resource heterogeneity, i.e., the distinction between persistent impoverished vs.\ rich environments, and thus produce qualitatively different results from those presented here.

The numerical experiments we discuss here were designed to explore
whether and how intrinsic mortality, a particularly extreme form of
individual restraint, might be evolutionarily favored in spatial
models. The results indeed showed that intrinsic mortality was
consistently favored, for a wide range of model assumptions and
parameters, with a very strong advantage over the lack of intrinsic
mortality.  Moreover, the results are consistent with a variety of
empirical observations, including some which have posed challenges for
traditional theories of aging and mortality.

\section*{Results}

\begin{figure}[tbp]
\begin{center}
\includegraphics[width=\columnwidth]{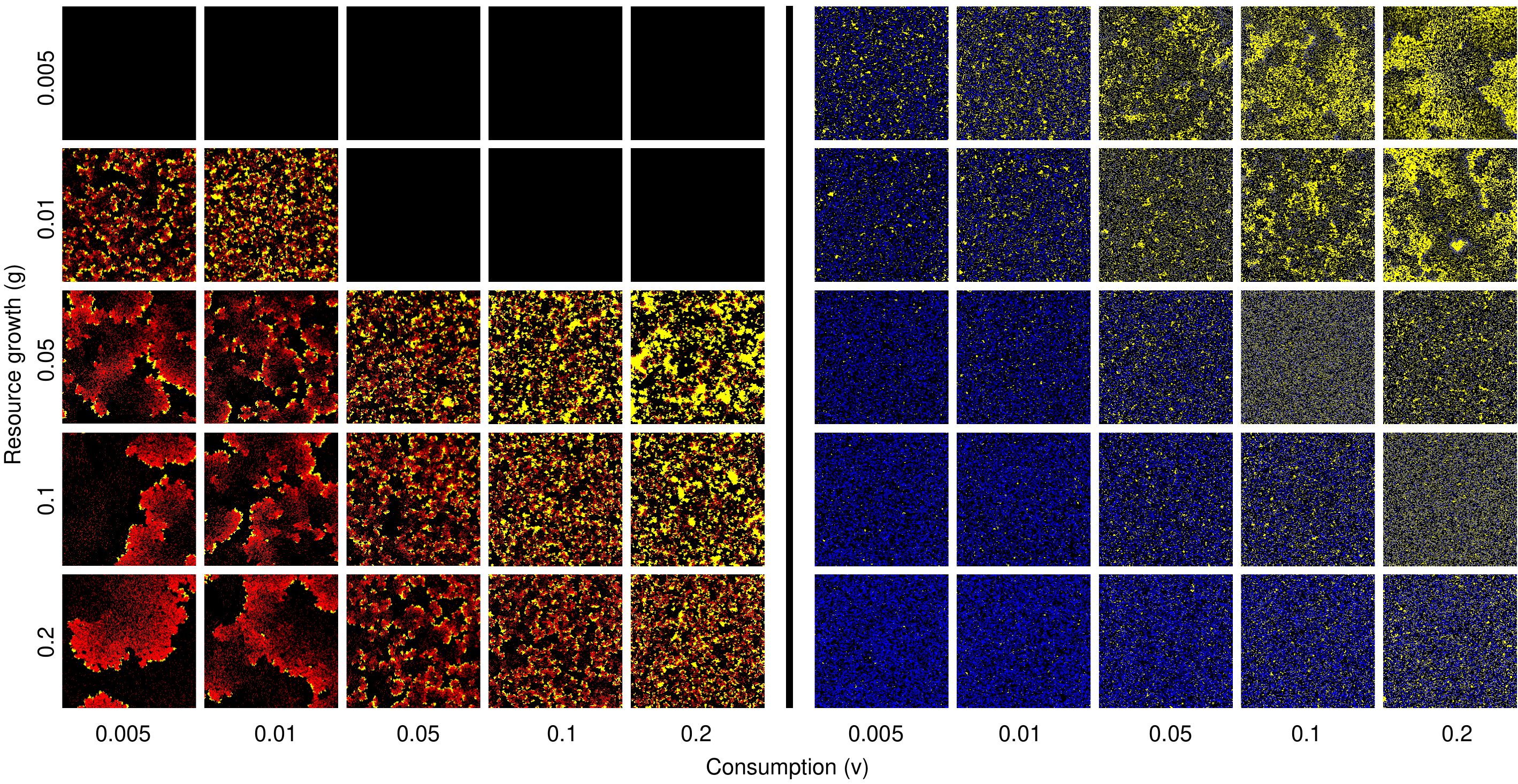}
\caption{{\bf Model snapshots showing different spatial distributions
    of consumers and resources.}  Resources are shown in yellow;
  consumers with intrinsic mortality (nonzero $q$) are shown in blue,
  those without it ($q$ fixed at 0) in red; empty spaces are shown in
  black.  The values of $p$ and $q$ for each panel are those which
  evolve in ascendance studies (Fig.~3B,C).  Each panel shows
  $250\times 250$ sites; parameter values for which populations
  without intrinsic mortality were not found to be stable are shown as
  empty lattices (details and full lattice sizes in the Models
  section).}
\label{snaps}
\end{center}
\end{figure}

We considered a family of explicitly spatial models of an interacting
pair of organism types, in which one population lives at the expense
of the second, as in predator-prey, pathogen-host, or herbivore-plant
systems.  Since analytic treatments \cite{prl15} average over or
otherwise approximate the complex dynamic patterns that are
responsible for the distinct behavior of spatial models, we relied
primarily on direct simulation to characterize their behavior.  The
model (Fig.~\ref{snaps}) used a stochastic cellular automaton to
consider a population of abstract organisms (``consumers'') with
limited but self-renewing resources (e.g., prey or hosts). Sites in a
two-dimensional lattice represented empty space, available resources,
or a consumer together with resources (a consumer could not occupy a
site in the absence of resources).  At each time step, resources alone
reproduced into neighboring empty sites, with probability $g$ for each
empty site; consumers reproduced into neighboring resource-only sites,
with probability $p$ for each resource site; consumers exhausted
resources in their site, leaving empty space, with probability $v$ per
time step and $c$ per reproduction; and consumers died due to
intrinsic mortality, leaving resources, with probability $q$.  A
consumer's mean intrinsic lifespan is $L=1/q$.  To explore the
evolution of lifespan in the consumer population, the intrinsic death
rate $q$ as well as the reproduction rate $p$ were heritable and
independently subject to mutation: with probability $\mu_q$ ($\mu_p$),
a consumer offspring has a value of $q$ ($p$) differing from that of
the parent by $\pm\epsilon_q$ ($\pm\epsilon_p$).

\begin{figure}[tbp]
\begin{center}
\includegraphics[width=.7\columnwidth]{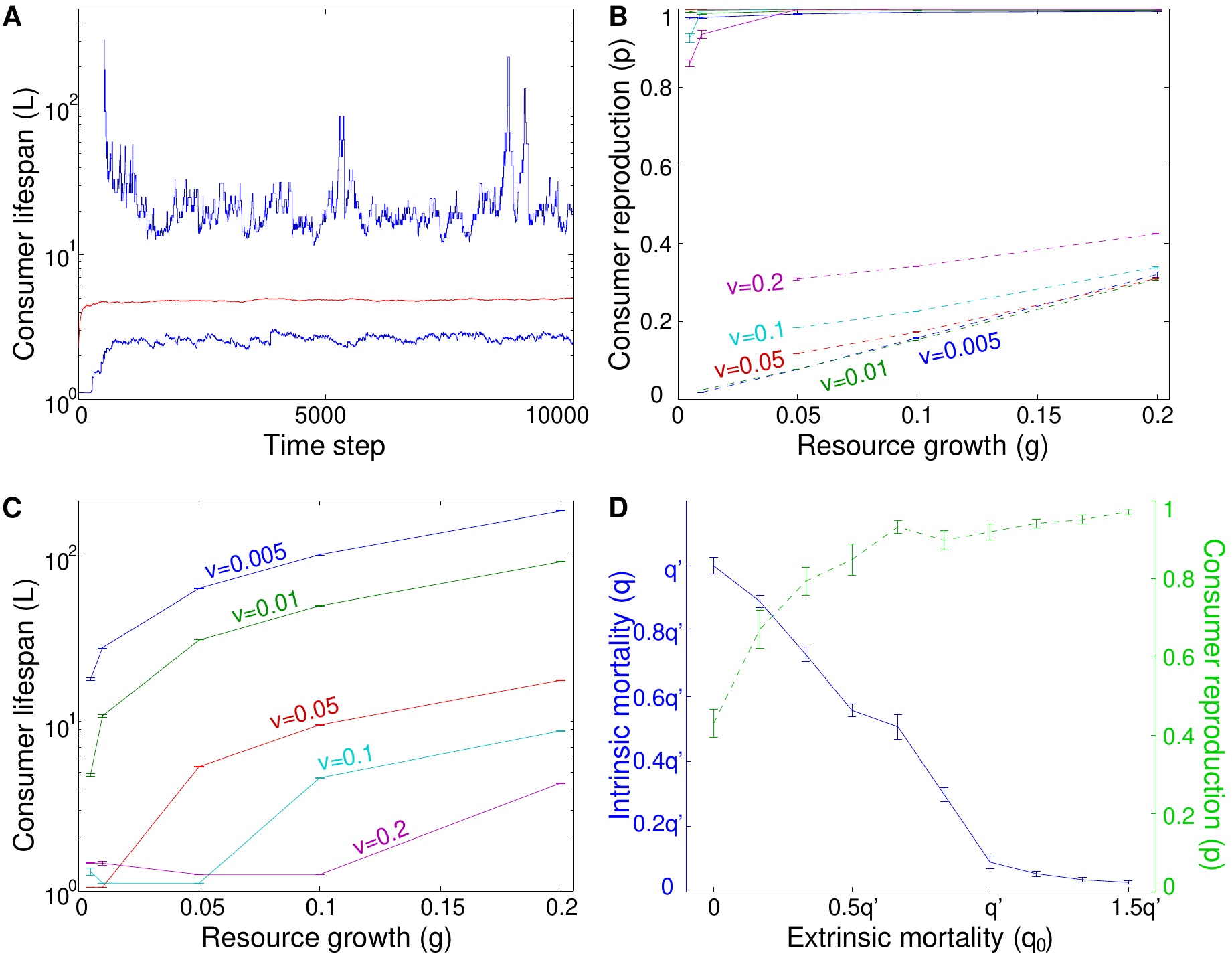}
\caption{{\bf Ascendance studies favor self-limited lifespan.} (A)
  History of evolving consumer lifespan in one example simulation
  ($g=v=0.1, c=0$), showing population mean/maximum/minimum. (B,C)
  Steady-state average values of (B) consumer reproduction probability
  $p$ and (C) intrinsic lifespan $L=1/q$, for different values of
  parameters $g$ and $v$ and for populations with (solid) and without
  (dashed) intrinsic mortality. (D) Adding an additional source of
  extrinsic mortality of magnitude $q_0$ decreases the evolved value
  of intrinsic mortality $q$, and increases evolved reproduction
  probability $p$. The unit of measure $q'$ is the mean value of $q$
  that evolves for $q_0=0; g=v=0.01, c=0.4$. All error bars show the
  standard error of the mean from ten independent trials.}
\label{fig:results}
\end{center}
\end{figure}

\begin{figure}[tbp]
\begin{center}
\includegraphics[width=.7\columnwidth]{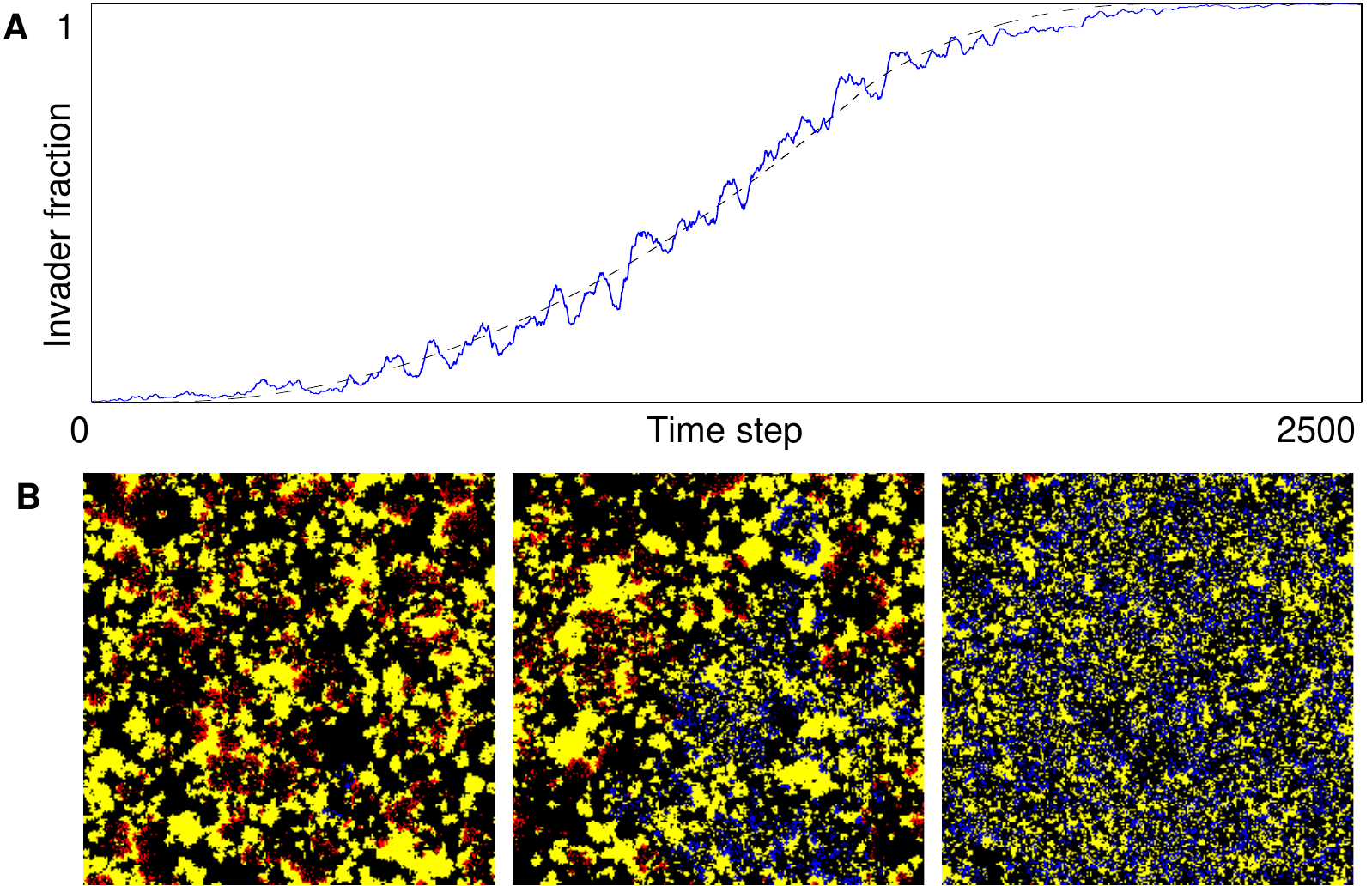}
\caption{{\bf A successful invasion of consumers without intrinsic mortality by those with the capacity for programmed death.} (A) The fraction of invaders in the population (solid line) increases almost monotonically with time. The region dominated by consumers with the capacity for programmed death grows steadily: the dashed line shows the area of a circle (under periodic boundary conditions) whose radius increases at a constant rate (correlation $r=0.997$). Resource growth $g=0.05$, consumer consumption $v=0.2$, consumer reproduction cost $c=0$. (B) Snapshots at 150, 1250, and 2350 time steps (colors as in Fig.~\ref{snaps}).}
\label{graphs}
\end{center}
\end{figure}

Two types of simulations investigated the evolution of lifespan
control in a population. In ``ascendance'' studies
(Fig.~\ref{fig:results}), we tracked the evolution of the intrinsic
death rate $q$ and reproduction rate $p$ over time in a randomly
initialized consumer population, to investigate what values dominate
in the long term.  In ``invasion'' studies (Fig.~\ref{graphs}), we
considered the ability of a single consumer to take over a population,
in order to ask: If a rare mutation could confer or remove the
capacity for lifespan control, would that mutant have an advantage or
a disadvantage in its later spread through the population? These
studies introduced one consumer into a steady-state population,
followed its lineage until fixation (extinction of either invaders or
invaded), and examined the probability of successful invasion in
100,000 or more such trials.

Additionally, in order to investigate the robustness of the results to model assumptions, we further performed ascendance studies with numerous variants of the model. These explored a variety of considerations relevant to the model's applicability to real-world systems, to test whether the results were sensitive to these factors. Variants included:
\begin{itemize}
\item explicit deterministic lifespan and
\item other age-dependent senescence patterns, rather than a fixed probability of death per time step;
\item local consumer mobility through rearrangement and/or migration;
\item increased dispersal range of consumers or resources;
\item resource depletion that is deterministic and gradual, rather than stochastic and binary;
\item ability of consumers to adjust their rate of living in response to resource availability;
\item spontaneous resource generation;
\item consumers reproducing to replace neighbors in occupied sites;
\item resources reproducing while exploited by consumers;
\item sexual reproduction by consumers.
\end{itemize}

The ascendance studies showed that self-limited lifespan was consistently favored, for all parameter values tested. Changing the values of the constant parameters $g$ and $v$ changed the population size and distribution of consumers and resources (Fig.~\ref{snaps}) and specific equilibrium value of lifespan that evolved (Fig.~\ref{fig:results}C); but the qualitative result, selection for lifespan limitation, was found in all cases.

The capacity for intrinsic mortality changed the characteristic
population structure, as qualitatively visible in Fig.~\ref{snaps}.
Table \ref{Table1} shows quantitatively that the mean patch size
(defined as the number of consumers in a connected set) was
consistently larger for populations without intrinsic mortality than
for those with it, for the same values of the ecological parameters
$\{g,v\}$.

\begin{table}[!htp]
\begin{center}
\caption{{\bf Intrinsic mortality changes characteristic population structure.}}
\begin{tabular}{|cc|c|c|c|c|c|}
\hline
&& $v=0.005$ & $v=0.01$ & $v=0.05$ & $v=0.1$ & $v=0.2$\\
\hline
\multirow{2}{*}{$g=0.005$} & Immortal & N/A & N/A & N/A & N/A & N/A\\
& Mortal & $8.7\pm 0.4$ & $7.2\pm 0.4$ & $1.01 \pm 0.01$ & $2\pm 1$ & $2.0\pm 0.1$\\
\hline
\multirow{2}{*}{$g=0.01$} & Immortal & $25\pm 1$ & $12.5\pm 0.3$ & N/A & N/A & N/A\\
& Mortal & $8.2\pm 0.3$ & $7.9\pm 0.3$ & $1.01\pm 0.01$ & $1.02\pm 0.01$ & $2.0\pm 0.5$\\
\hline
\multirow{2}{*}{$g=0.05$} & Immortal & $250\pm 50$ & $79\pm 8$ & $12.3\pm 0.5$ & $8.2\pm 0.7$ & $6.3\pm 0.4$\\
& Mortal & $11.8\pm 0.4$ & $11.2\pm 0.6$ & $7.8\pm 0.4$ & $1.08\pm 0.03$ & $1.03\pm 0.01$\\
\hline
\multirow{2}{*}{$g=0.1$} & Immortal & $700\pm 300$ & $230\pm 50$ & $24\pm 1$ & $11.6\pm 0.9$ & $7.4\pm 0.7$\\
& Mortal & $20\pm 1$ & $18\pm 1$ & $12.2\pm 0.5$ & $8.1\pm 0.3$ & $1.07\pm 0.03$\\
\hline
\multirow{2}{*}{$g=0.2$} & Immortal & $5000\pm 4000$ & $700\pm 300$ & $50\pm 4$ & $21\pm 2$ & $10.5\pm 0.6$\\
& Mortal & $54\pm 5$ & $46\pm 6$ & $28\pm 2$ & $17\pm 1$ & $8.5\pm 0.2$\\
\hline
\end{tabular}
\begin{flushleft}  Mean consumer patch size, averaged over all consumers, for different values of resource growth rate $g$ and consumption rate $v$, and for populations with and without intrinsic mortality (``mortal'' and ``immortal'', respectively).
\end{flushleft}
\label{Table1}
\end{center}
\end{table}

Adding extrinsic mortality (corresponding to deaths through, e.g., predation or accidents) reduced the evolved level of intrinsic mortality $q$, such that the total mortality level favored by selection for a given set of parameters remained approximately constant (Fig.~\ref{fig:results}D).

The model variants showed that the key finding of selection for lifespan limitation was robust to changes in model details and assumptions. In almost every case, the outcome was the evolution of self-limited lifespan. The exceptions were consistent with this understanding and with previous work: When consumer dispersal was large compared to the size of the space, the system was effectively well-mixed and lifespan control did not evolve. When resource availability was so great as to support the maximum amount of consumption and reproduction the model allowed, the system evolved to these limits. In real-world systems, both dispersal and resources are normally limited.

The invasion studies showed a very strong population advantage to the capacity for intrinsic mortality (Table \ref{Table2}). Consumers with this capacity ($q=0$ for the invader, but potentially nonzero for its descendants through mutation) invading populations of consumers without intrinsic mortality ($q$ fixed at 0 for all descendants) had a success rate typically two to three orders of magnitude greater than that of invaders without intrinsic mortality, across all values of the ecological parameters $g$ and $v$ tested. Conversely, consumers with no intrinsic mortality managed no successful invasions of populations with programmed death in a total of several million trials.

\begin{table}[tbp]
\begin{center}
\caption{{\bf Consumers with the capacity for intrinsic mortality dominate those without it in invasion experiments.}}
\begin{tabular}{cc|c|c|}
\cline{3-4}
\multicolumn{2}{c|}{\multirow{2}{*}{$g=0.05,v=0.05$}}
& \multicolumn{2}{|c|}{Invaded} \\
\cline{3-4}
\multicolumn{2}{c|}{}
& Immortal & Mortal \\
\hline
\multicolumn{1}{|c|}{\multirow{2}{*}{Invader}}
& Mortal & $(2.51\pm 0.05)\times 10^{-2}$ & $(1.0\pm 0.7)\times 10^{-5}$ \\
\cline{2-4}
\multicolumn{1}{|c|}{}
& Immortal & $(2\pm 1)\times 10^{-5}$ & $0$ \\
\hline
\end{tabular}
\vspace*{.5cm}

\begin{tabular}{cc|c|c|}
\cline{3-4}
\multicolumn{2}{c|}{\multirow{2}{*}{$g=0.05,v=0.1$}}
& \multicolumn{2}{|c|}{Invaded} \\
\cline{3-4}
\multicolumn{2}{c|}{}
& Immortal & Mortal \\
\hline
\multicolumn{1}{|c|}{\multirow{2}{*}{Invader}}
& Mortal & $(1.91\pm 0.04)\times 10^{-2}$ & $(1.1\pm 0.3)\times 10^{-4}$ \\
\cline{2-4}
\multicolumn{1}{|c|}{}
& Immortal & $(1.2\pm 0.3)\times 10^{-4}$ & $0$ \\
\hline
\end{tabular}
\vspace*{.5cm}

\begin{tabular}{cc|c|c|}
\cline{3-4}
\multicolumn{2}{c|}{\multirow{2}{*}{$g=0.05,v=0.2$}}
& \multicolumn{2}{|c|}{Invaded} \\
\cline{3-4}
\multicolumn{2}{c|}{}
& Immortal & Mortal \\
\hline
\multicolumn{1}{|c|}{\multirow{2}{*}{Invader}}
& Mortal & $(1.23\pm 0.04)\times 10^{-2}$ & $(1.6\pm 0.4)\times 10^{-4}$ \\
\cline{2-4}
\multicolumn{1}{|c|}{}
& Immortal & $(2.8\pm 0.5)\times 10^{-4}$ & $0$ \\
\hline
\end{tabular}
\vspace*{.5cm}

\begin{tabular}{cc|c|c|}
\cline{3-4}
\multicolumn{2}{c|}{\multirow{2}{*}{$g=0.1,v=0.05$}}
& \multicolumn{2}{|c|}{Invaded} \\
\cline{3-4}
\multicolumn{2}{c|}{}
& Immortal & Mortal \\
\hline
\multicolumn{1}{|c|}{\multirow{2}{*}{Invader}}
& Mortal & $(2.51\pm 0.05)\times 10^{-2}$ & $(2\pm 1)\times 10^{-5}$ \\
\cline{2-4}
\multicolumn{1}{|c|}{}
& Immortal & $(2\pm 1)\times 10^{-5}$ & $0$ \\
\hline
\end{tabular}
\vspace*{.5cm}

\begin{tabular}{cc|c|c|}
\cline{3-4}
\multicolumn{2}{c|}{\multirow{2}{*}{$g=0.1,v=0.1$}}
& \multicolumn{2}{|c|}{Invaded} \\
\cline{3-4}
\multicolumn{2}{c|}{}
& Immortal & Mortal \\
\hline
\multicolumn{1}{|c|}{\multirow{2}{*}{Invader}}
& Mortal & $(2.38\pm 0.05)\times 10^{-2}$ & $(5\pm 2)\times 10^{-5}$ \\
\cline{2-4}
\multicolumn{1}{|c|}{}
& Immortal & $(1.4\pm 0.4)\times 10^{-4}$ & $0$ \\
\hline
\end{tabular}
\vspace*{.5cm}

\begin{tabular}{cc|c|c|}
\cline{3-4}
\multicolumn{2}{c|}{\multirow{2}{*}{$g=0.1,v=0.2$}}
& \multicolumn{2}{|c|}{Invaded} \\
\cline{3-4}
\multicolumn{2}{c|}{}
& Immortal & Mortal \\
\hline
\multicolumn{1}{|c|}{\multirow{2}{*}{Invader}}
& Mortal & $(1.80\pm 0.04)\times 10^{-2}$ & $(1.1\pm 0.3)\times 10^{-4}$ \\
\cline{2-4}
\multicolumn{1}{|c|}{}
& Immortal & $(1.1\pm 0.3)\times 10^{-4}$ & $0$ \\
\hline
\end{tabular}
\vspace*{.5cm}

\begin{tabular}{cc|c|c|}
\cline{3-4}
\multicolumn{2}{c|}{\multirow{2}{*}{$g=0.2,v=0.05$}}
& \multicolumn{2}{|c|}{Invaded} \\
\cline{3-4}
\multicolumn{2}{c|}{}
& Immortal & Mortal \\
\hline
\multicolumn{1}{|c|}{\multirow{2}{*}{Invader}}
& Mortal & $(2.24\pm 0.05)\times 10^{-2}$ & $(2\pm 1)\times 10^{-5}$ \\
\cline{2-4}
\multicolumn{1}{|c|}{}
& Immortal & $(3\pm 2)\times 10^{-5}$ & $0$ \\
\hline
\end{tabular}
\vspace*{.5cm}

\begin{tabular}{cc|c|c|}
\cline{3-4}
\multicolumn{2}{c|}{\multirow{2}{*}{$g=0.2,v=0.1$}}
& \multicolumn{2}{|c|}{Invaded} \\
\cline{3-4}
\multicolumn{2}{c|}{}
& Immortal & Mortal \\
\hline
\multicolumn{1}{|c|}{\multirow{2}{*}{Invader}}
& Mortal & $(2.44\pm 0.05)\times 10^{-2}$ & $(5\pm 2)\times 10^{-5}$ \\
\cline{2-4}
\multicolumn{1}{|c|}{}
& Immortal & $(1.1\pm 0.3)\times 10^{-4}$ & $0$ \\
\hline
\end{tabular}
\vspace*{.5cm}

\begin{tabular}{cc|c|c|}
\cline{3-4}
\multicolumn{2}{c|}{\multirow{2}{*}{$g=0.2,v=0.2$}}
& \multicolumn{2}{|c|}{Invaded} \\
\cline{3-4}
\multicolumn{2}{c|}{}
& Immortal & Mortal \\
\hline
\multicolumn{1}{|c|}{\multirow{2}{*}{Invader}}
& Mortal & $(2.12\pm 0.5)\times 10^{-2}$ & $(1\pm 1)\times 10^{-5}$ \\
\cline{2-4}
\multicolumn{1}{|c|}{}
& Immortal & $(7\pm 3)\times 10^{-5}$ & $0$ \\
\hline
\end{tabular}
\begin{flushleft} Probabilities of successful invasions for different values of
    resource growth rate \emph{g} and consumption rate \emph{v} (all
    with reproduction cost \emph{c}=0), and for each combination of
    invaders and invaded having or lacking the capacity for intrinsic mortality (``mortal'' and ``immortal'', respectively).
\end{flushleft}
\label{Table2}
\end{center}
\end{table}

\section*{Discussion}

The simulation results provide theoretical support for the idea that
direct lifespan control, and programmed mortality and senescence as
ways of achieving it, are consistent with natural selection. In
contrast, it is widely reported that theory is incompatible with the
evolution of explicit lifespan control
\cite{KirkwoodAustad00,OlshanskyEtAl02,Kirkwood05}.  This perspective
has guided and constrained the interpretation of empirical findings.
Our results suggest that classic mechanisms relevant to the evolution
of lifespan are incomplete.  The mutation accumulation and
antagonistic pleiotropy theories provide established mechanisms
contributing to gradual senescence; direct selection provides an
additional mechanism, not a replacement---where multiple mechanisms
are possible in biology, typically all may play a role. Our results
suggest the broad applicability of an additional mechanism missing
from previous theories.  This mechanism can provide additional
interpretations and novel understanding of a variety of observations
in nature.

\subsection*{Empirical phenomena: providing needed explanations}

Direct selection for intrinsic mortality and senescence---not just selection for an individual benefit with senescence as a side effect---can be used to help understand empirical phenomena, particularly for cases that have posed problems for traditional theory \cite{Kirkwood05} but are straightforward to explain with direct selection. Here we briefly discuss three such cases.

(1) Organisms of some species reproduce only once and die after
reproducing (semelparity), and may live for a variable period
\cite{Finch90} before the act of reproduction apparently triggers
their death. Because programmed death is inconsistent with classic
theories \cite{Kirkwood05}, other interpretations for semelparity have
been sought; the traditional explanation is that so much energy goes
into reproduction that none is left over for essential physiological
maintenance afterwards \cite{Kirkwood05}. However, cases such as
octopus where removal of the optic gland allows continued survival and
resumption of copulation \cite{Wodinsky77} are difficult to reconcile
with this interpretation, and are simpler to explain with an explicit
programmed death mechanism.

(2) It is widely reported as a key prediction of traditional theories
that higher extrinsic mortality rates should result in the evolution
of shorter intrinsic lifespans
\cite{Williams57,KellerGenoud97,Ricklefs98,KirkwoodAustad00,Reznick01,Reznick04,Kirkwood05};
others hold that extrinsic mortality should have no effect on evolved
lifespan \cite{Caswell07}.  In contrast to both, a key empirical study
observed that guppies that evolved subject to higher levels of
predation exhibited longer lifespans and lower rates of aging
\cite{Reznick04}. The spatial model predicts this result because
additional extrinsic mortality reduces the need for intrinsic
mortality \cite{Goss94}.  Fig.~\ref{fig:results}D shows quantitatively
that adding a source of extrinsic mortality unrelated to resource
exhaustion (with probability $q_0$), by replacing the intrinsic death
probability $q$ with ($q+q_0$), leads to lower evolved intrinsic
mortality.

(3) It has been argued that in multicellular organisms---where the
distinctions between gametes and other cells (germline versus soma),
and between parent and offspring, are clear---senescence is
inevitable, while unicellular organisms (lacking those distinctions)
should be effectively immortal
\cite{Weismann1891,Williams57,KirkwoodAustad00,Kirkwood05,Evan94}. However,
certain multicellular organisms have been reported to have negligible
or even negative senescence, i.e., displaying insignificant or
declining effects of aging according to population-statistical or
physiological measures
\cite{Stiven62,Finch90,Ger94,Muller96,Ger94,Martinez98,GirandotGarcia98,ExpGerontol01Issue,VaupelEtAl04,Guerin04,Vaupel10}.
Conversely, senescence has been reported in organisms with no clear
germ/soma distinction \cite{MartinezLevinton92}, and programmed cell
death reported in unicellular organisms
\cite{Ameisen96,Lewis00,DuszenkoEtAl06,KolodkinGalEtAl07}. The
difficulty of reconciling such empirical observations with traditional
theories has been pointed out
\cite{MartinezLevinton92,Ameisen96,Martinez98,Lewis00,JamesGreen02,BaudischVaupel12}. These
phenomena are straightforward to explain if mortality and senescence
are determined by selection, with favored lifespan set according to
ecological conditions regardless of whether organisms are
single-celled or multicellular.  The spatial model predicts that
negligible senescence may be favored under appropriate ecological
conditions (e.g., well-mixed populations, or those kept in check
entirely by factors other than resource limitations); conversely,
programmed death can be favored for unicellular consumers as easily as
for multicellular ones.

We note that this potential uncoupling of intrinsic mortality and
multicellularity has implications for our understanding of
evolutionary history.  Programmed cell death, critical during
development in multicellular organisms (morphogenesis), has
traditionally been viewed as having multicellularity as a
prerequisite---the sharp distinction between somatic and germ cells
intuitively enables some of the former to die without compromising
their genetic contribution to the next generation---and accordingly it
has been argued to have arisen contemporaneously or later
\cite{Evan94,VauxEtAl94,AravindEtAl01}. Our results suggest that
programmed cell death likely arose first in unicellular populations
\cite{Shapiro98,Ameisen02,JamesGreen02,KooninAravind02,GrosbergStrathmann07}
and later become co-opted for morphogenesis, perhaps helping to enable
the evolutionary appearance of multicellularity.  Similarly, the
reproductive senescence observed in some unicellular organisms
\cite{Finch90} could have been an exaptation for the cessation of
division in somatic cell lines.

\subsection*{Empirical phenomena: providing alternate interpretations}

Finding lifespan control in a robust theoretical model opens the door
to interpreting a wide range of observations of aging-related
phenomena as arising from natural selection. Such a direct-selection
mechanism does not preclude classic mechanisms such as mutation
accumulation and antagonistic pleiotropy, but rather augments them,
providing additional explanations that may contribute to observed
phenomena.  Here we suggest examples of cases where future study may
find that selection directly for lifespan control plays a significant
role. Since the model conditions essential to the qualitative result
of self-limited lifespan---spatial environment, exhaustible/renewable
resource, and limited dispersal---are typical features of the real
world, we would expect lifespan control to be a widely present
phenomenon.

(1) The great range of natural lifespans observed among organisms, both dissimilar ones and those otherwise similar \cite{Finch90,Love02}, is traditionally explained as selection for longer life being insufficiently strong in the cases of shorter-lived organisms to extend their lifespans to match those of longer-lived ones. Instead or additionally, shorter lifespans could be actively selected for in some organisms, according to ecological conditions.

(2) Large single-gene or few-gene effects on lifespan \cite{LakowskiHekimi96} were initially surprising from the viewpoint of traditional theories \cite{Kirkwood05}.  Some have been interpreted as examples of antagonistic pleiotropy \cite{Williams57}, where increased lifespan carries a fitness cost in early life \cite{Walker00}. A more direct interpretation is that they could represent effective means of control to regulate the tuning of lifespan for different circumstances.

(3) The evolutionary loss of mouthpart function in many adult insects, such as mayflies, is often the primary factor limiting lifespan \cite{Finch90}. The traditional explanation for this loss of function is evolutionary irrelevance after reproduction; instead, it could have come about through active selection as a means of limiting lifespan. Similar mechanisms may be present in other animals: for instance, elephants go through six sets of molars in their lifetime, and starvation after the final set wears out may be the leading cause of death for older elephants \cite{Finch90}.

(4) The model predicts that reproduction rate rises with increasing extrinsic mortality rate (Fig.~\ref{fig:results}D). This is consistent with a field study of an island population of opossums, free from predators, which were observed to live longer and reproduce more slowly than mainland individuals \cite{Austad93}. These trends have been interpreted as evidence of increased senescence in the presence of increased predation \cite{Austad93,Miller02}. However, an alternate interpretation is that the shorter observed \emph{natural} lifespan on the mainland is a direct result of the high predation (\emph{intrinsic} lifespan is rarely observed there, as most opossums are killed by predators, with barely a quarter of females surviving to produce a second litter \cite{Austad93}), and the reproduction rate is higher to compensate.

(5) The model predicts the evolution of long lifespans for low consumption $v$ relative to resource availability (Fig.~\ref{fig:results}C, \cite{prl15}). This is consistent with observations of animals such as marine fishes, which have longer lifespans and slower metabolisms at greater depths \cite{CaillietEtAl01}; crocodilians, which go from weeks to years between meals \cite{Ross89} and are long-lived species characterized as exhibiting negligible senescence \cite{Patnaik94}; and cave animals, which, compared to surface dwellers, appear to have adapted to conditions of food scarcity by evolving significantly extended lifespan \cite{CulverPipan}. Similarly, populations limited by factors other than resource availability are predicted to evolve longer lifespans (e.g.,
as with the guppies subject to higher predation and having longer intrinsic lifespans \cite{Reznick04}).

(6) The model demonstrates that in many circumstances, a strain exhibiting increased reproduction along with increased death rate has an advantage over slower-reproducing, longer-lived individuals. This is consistent with studies of cancer where rapidly replicating cells also have increased rates of apoptosis \cite{TormanenEtAl99}, suggesting an advantage over other possible cancerous types as well as normal cell populations.

(7) It is sometimes raised as an objection to hypotheses of programmed aging that while many genes have been found which have large effects on lifespan, none has been found which eliminates aging altogether \cite{Kirkwood05}. However, if self-limited lifespan has as powerful an adaptive advantage under as general conditions as our results suggest, it is to be expected that the genetic basis for it would be solidly and robustly established, with mechanisms for flexibly tuning lifespan to adjust to changing conditions but no simple way of eliminating the control altogether.

(8) The model demonstrates an evolutionary advantage associated with reducing resource use through shortening lifespans. Other mechanisms for reducing resource use (particularly age-dependent) may have similar effects. For instance, metabolism typically declines with age \cite{Kara94}, which may allow greater longevity while leaving more resources for descendants. A related issue is that of post-reproductive care of indirect offspring in humans (the ``grandmother hypothesis''). A strong relationship exists between post-reproductive lifespan in women and their number of grandchildren \cite{LahdenperaEtAl04}. (Note that a form of time-dependent fitness, measuring reproductive output at later generations, is necessary to describe this effect: there is no relationship between a woman's lifespan and her number of direct offspring \cite{LahdenperaEtAl04}.) The benefits associated with grandmother care may compete against the benefits of increased resource availability through early death that we have focused on. In organisms where such care is present, its effects oppose selection for shorter lifespan, with potential relevance to the notably long lifespans of humans and perhaps some other primates \cite{BronikowskiEtAl11,JudgeCarey00,WalkerHerndon08}.

\subsection*{Time-dependent fitness}

\begin{figure}[tbp]
\begin{center}
\includegraphics[width=0.5\textwidth]{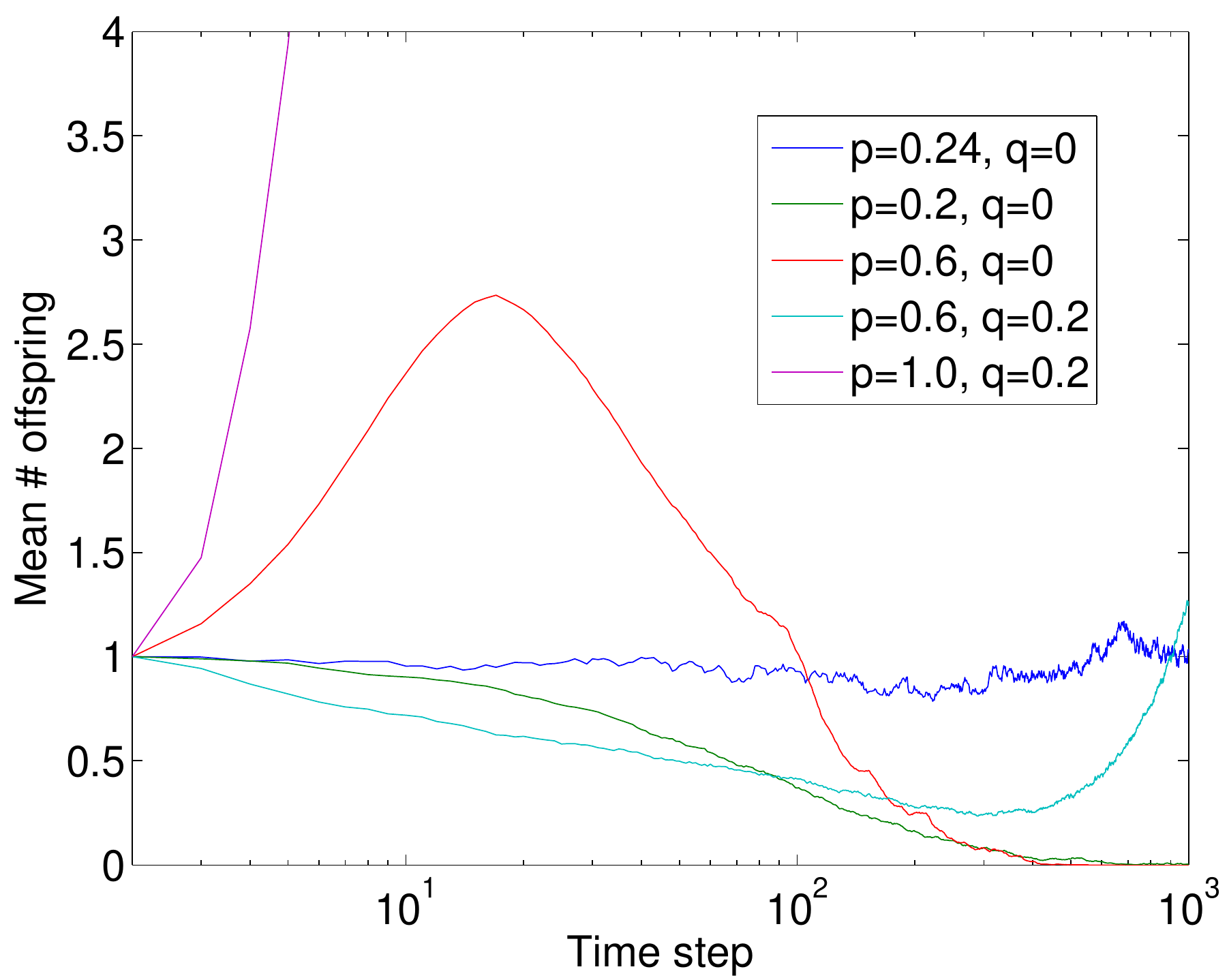}
\caption{Mean number of descendants of a consumer with specified $p$
  and $q$, introduced into a steady-state population with $g=0.1,
  v=0.1, p=0.24, q=0, \mu_{p}=\mu_{q}=0$.  Without intrinsic mortality
  ($q=0$), a higher reproduction probability gives a short-term
  advantage to a lineage, but on longer timescales the average number
  of surviving descendants plummets.  Intrinsic mortality allows
  faster reproduction, giving a long-term advantage despite a
  disadvantage that may last hundreds or thousands of time steps
  ($p=0.6, q=0.2$), or even favoring a lineage on all time scales
  ($p=1.0, q=0.2$).}
\label{timedepfit}
\end{center}
\end{figure}

A useful characterization of the long-term behavior of evolutionary
models is the average number of descendants of a given type after an
amount of time $T$---the time-dependent fitness
\cite{RauchEtAlPRL02,RauchEtAlJTB03}. This measure quantifies the
long-term survival of the type including effects of changes caused by
the type to its environment. Simulations to obtain the time-dependent
fitness are similar to invasion studies, but without
mutation. Fig.~\ref{timedep} showed results within populations of
mortal consumers.  Examples of results within a population of immortal
consumers are shown in Fig.~\ref{timedepfit}: Immortal consumers with
$p$ at its evolutionarily stable value (here, 0.24) have a persistent
fitness approximately equal to 1. Higher reproduction probabilities
(e.g., $p=0.6$) give a significant short-term advantage, for hundreds
of time steps, but subsequently the average number of surviving
descendants falls to 0. Mortal consumers (with faster reproduction,
but programmed death) can have the opposite pattern, with a short-term
disadvantage, but a significant advantage in the long term (e.g.,
$p=0.6, q=0.2$).

     This result illustrates that immediate advantage or disadvantage does not necessarily determine long-term advantage---rather, processes operating over hundreds or thousands of generations can overcome more immediate considerations. Thus, kin selection alone \cite{Bourke07} is insufficient to account for these phenomena of restraint (self-limited reproduction or lifespan): Kin selection is concerned with same-generation interactions, and is sometimes extended to immediate descendants; to characterize kin selection as relevant to descendants arbitrarily far in the future is to imply that the kin-selectionist argument is equivalent to evolution itself, in which case the concept provides no additional conceptual or mathematical benefit.

     In other cases, mortal consumers can have an advantage over immortals in both the short and long term (e.g., $p=1.0, q=0.2$).

Looking at number of offspring as a function of generation number rather than time (Fig.~\ref{timedep}, right-hand plots) provides a connection to population growth rate in traditional theoretical treatments. If $x$ is the generation number and $y(x)$ is the number of invaders of that generation (y-value in the plot), then $y(1)$ is the average number of direct offspring of an initial invader; similarly, $y(x+1)/y(x)$ gives the average lifetime reproduction per consumer in later generations.  Dividing this quantity by the consumer lifetime, $1/(v+kc+q)$ (taking all causes of mortality into account), gives $r$, the average number of offspring per consumer per time step. The population growth rate is this birth rate minus the death rate, $r-(v+kc+q)$. However, as those plots show, an initial advantage in population growth rate does not necessarily characterize long-term reproductive success.  The traditional single-generation measures of fitness and population growth rate are not sufficient descriptors of long-term success in spatial systems.

\subsection*{Classic perspectives on lifespan control}

Mainstream evolutionary theory considers it well-established that
selection does not and cannot act in favor of decreased lifespan
(unless more than compensated by a concomitant increase in early-life
fitness, in the antagonistic pleiotropy framework \cite{Williams57}).
A few characteristic quotations illustrate this point:

\begin{itemize}

\item ``The way evolution works makes it impossible for us to possess genes that are specifically designed to cause physiological decline with age or to control how long we live. Just as an automobile does not have a built-in plan for decline written in its blueprints, we do not possess genetic instructions that tell our bodies how to age or when to die.'' \cite{OlshanskyEtAl02}

\item ``Evolutionary theory correctly asserts that aging is not an adaptive trait.'' \cite{HekimiGuarente03}

\item ``There is a widespread but erroneous tendency to regard aging as programmed.\ldots There are powerful arguments why [such a program] should not exist.\ldots The suggested benefits from aging are ones that serve the interests of the species or group. Whenever a benefit at the \emph{group} level is assumed to supersede the contrary interests of the \emph{individual}, any evolutionary hypothesis must confront the problem of `cheating.' Individuals in whom the aging program was inactivated by mutation would benefit from the sacrifice of others, while enjoying any fitness advantage that might accrue from immortality. Such mutations would therefore be expected to spread.'' \cite{Kirkwood05}

\item ``Any hypothetical `accelerated ageing gene' would be disadvantageous to the individual. It is therefore difficult to see how genes for accelerated ageing could be maintained in stable equilibrium, as individuals in whom the genes were inactivated by mutation would enjoy a selection advantage.'' \cite{KirkwoodAustad00}

\item ``Aging, a decline in condition with increasing age apparent as a reduction in survival and reproductive output, is apparently disadvantageous for the individual. Why then, does it exist at all? Is there some hidden advantage to aging? Evolutionary biologists tackled these questions early on and concluded that aging does not have a function and exists only because natural selection is less powerful late in life.\ldots These two basic tenets, that aging is due to a declining force of natural selection and is not adaptive, still form the conceptual foundation of the biology of aging.'' \cite{PletcherEtAl07}

\item ``There is a striking discrepancy between the diversity of theory on the evolution of senescence and its treatment in the literature. Empirical evaluations of the evolution of senescence focus almost exclusively on the classical theory, as do recent reviews.'' \cite{Reznick04}
\end{itemize}

These views can be traced back to Williams, who, in the article in which he proposed antagonistic pleiotropy, wrote:

\begin{itemize}
\item ``Natural selection should ordinarily proceed towards lengthening life, not shortening it. Such selection, at the individual level, could conceivably be countered by selection at the population level, if senescence somehow favored group-survival.\ldots The efficacy of such selection depends upon a rather complicated series of assumptions \ldots A theory based on the simpler and more widely applicable principle of selection within a group would be preferable, unless the assumption of effective between-group selection proves to be necessary.'' \cite{Williams57}
\end{itemize}

In this way Williams argued that selection for shorter lifespan should not be considered if any alternative explanation exists. Such a view risks blinding evolutionary biology to an important explanatory process, particularly with increasing evidence that selection above the individual level is an important evolutionary force\cite{RauchEtAlPRL02,RauchEtAlJTB03,pnas04,WilsonHolldobler05,WilsonWilson07,Wilson08,NowakTarnitaWilson10}.

\subsection*{Context and future work}

Previous models that predict selection for increased mortality account for it through intrinsic physiological tradeoffs: e.g., by putting energy into reproduction at the expense of maintenance, an individual may achieve earlier reproductive increases that more than compensate for lost later opportunities.  In our model, the tradeoffs that ultimately result in greater net reproduction are imposed through an extrinsic mechanism of environmental feedback: dying earlier can leave resources that can most effectively be used by descendants, with the net reproductive benefits appearing generations later.

The idea that shorter lifespans can be and are selected for directly
goes back to at least 1870 \cite{Weismann1891}. It was later rejected
based on theoretical arguments that evolution of such a trait opposed
to individual self-interest, like other altruistic behaviors, must
require group selection, whose applicability should be accepted only
as a last resort \cite{Williams57}.  Analytic \cite{prl15} and
experimental \cite{StearnsEtAl00} studies that take a mean-field or
spatially mixed approach support the conclusions of the now-standard
theories.  However, such spatially averaged systems exhibit
qualitatively different behavior from systems which, like the real
world, possess spatial extent. In particular, spatial systems
routinely demonstrate altruistic behaviors
\cite{RauchEtAlPRL02,RauchEtAlJTB03,pnas04,Nowak06,AustinEtAl08} which
are not evolutionarily stable in spatially averaged (mean-field)
models \cite{BarYam97,BarYam99,prl15} or well-mixed laboratory
populations \cite{AustinEtAl08}. Previous models, some spatial, have
demonstrated that selection for self-limited lifespan is not a
theoretical impossibility \cite{LonerganTravis03,Leslie06}, under
assumptions such as continual introduction of highly advantageous
mutations \cite{Libertini88}, pre-existing senescence in the form of
decreasing fecundity \cite{Travis04,DythamTravis06} or decreasing
competitive fitness \cite{Martins11} with increasing age, frequent
deadly epidemics among physically and genetically close relatives
\cite{MitteldorfPepper09}, or explicit group selection among
nearly-isolated subpopulations \cite{Mitteldorf06}. However, a
generally applicable mechanism for the active selection of lifespan
control without such limiting assumptions has not been previously
demonstrated.

The robustness of our result that self-limited lifespan is favored,
under many variations of model details and parameter values, suggests
that selection in favor of shorter lifespan and genetically programmed
senescence may indeed be a quite general phenomenon. As such, it may
have acted on the ancestors of human beings, with strong implications
for human medicine.  If aging is a functional adaptation, rather than
a collection of inevitable breakdowns or genetic tradeoffs, then
effective health and life extensions through dietary, pharmacological,
or genetic interventions
\cite{FinkelHolbrook00,MartinOshima00,Hayflick00,Harris00,DeGrey02,Miller02,Goldsmith08,BakerEtAl11,SinhaEtAl14}
are likely to be possible, with potential for significant impact
(e.g., altering two genes extends nematode lifespan fivefold
\cite{LakowskiHekimi96}).  The effects of aging may one day be treated
through manipulation of an underlying mechanism rather than as
disparate symptoms.  The fact that theoretical understanding of
evolution can play a critical role in guiding health research should
motivate a wider reevaluation of the evidence in relation to the
theoretical frameworks.

\section*{Models}

In the base model whose results are plotted in the figures, the
dynamics comprise a sequence of synchronous updates of a
two-dimensional spatial array of cells. At each time step, the
following events simultaneously take place for each site in the array:

\begin{itemize}

\item An empty site has probability
$$
P_{E\rightarrow R} = 1-(1-g)^{\mathcal{N}_R}
$$
of transitioning to a resource-only site, where $g$ is the probability per time step of a resource-only site reproducing into a given neighboring empty site, and $\mathcal{N}_R$ is the number of resource-only sites among the empty site's four nearest neighbors. This expression is equivalent to the statement that each neighboring resource-only site has an independent probability $g$ of ``seeding'' the empty site.

\item A resource-only site has probability
$$
P_{R\rightarrow C} = 1-\prod_i^{\mathcal{N}_C}(1-p_i)
$$
of transitioning to a consumer site, where $\mathcal{N}_C$ is the number of consumers among its four neighbors and the values $p_i$ are the corresponding consumer reproduction probabilities. This expression corresponds to each neighboring consumer site $i$ having an independent probability $p_i$ of trying to reproduce into this site. If more than one consumer does so, one is chosen at random to be the parent. The offspring has $p$ and $q$ initialized equal to that of its parent; then with probability $\mu_p$ ($\mu_q$), that value is increased by $\epsilon_{p}$ ($\epsilon_q$), or decreased by that amount with equal probability.

\item A consumer site has probability $P_{C\rightarrow E} = v+kc$ of transitioning to an empty site (all resources in the site are consumed, resulting in consumer death), where $k$ is the number of offspring the consumer produces in this time step, and probability $P_{C\rightarrow R} = q$ of transitioning to a resource-only site (death due to intrinsic mortality).
\end{itemize}
Once every 100 time steps, the maximum, minimum, and mean values of $p$ and $q$ in the consumer population are recorded.

Simulations were performed on lattices of size $250\times 250$ unless the consumer population was not stable (quickly going to extinction in such cases), particularly a problem for immortal populations. Accordingly, some results shown in Fig.~\ref{fig:results} were run on larger arrays, each on the smallest of \{$250\times 250$, $500\times 500$, $750\times 750$\} for which a steady-state population of consumers could persist. Increasing the lattice size further does not change the steady-state values of $p$ and $q$. For immortal consumers, we used:
\begin{itemize}
\item $250\times 250$: $g=\{0.05, 0.1, 0.2\}, v=\{0.1, 0.2\}$
\item $500\times 500$: $g=\{0.05, 0.1\}, v=\{0.005, 0.01, 0.05\}$; $g=0.2, v=\{0.01, 0.05\}$
\item $750\times 750$: $g=0.01, v=\{0.005, 0.01\}$; $g=0.2, v=0.005$
\end{itemize}
In other cases ($g=0.005, v=\{0.005, 0.01, 0.05, 0.1, 0.2\}; g=0.01, v=\{0.05, 0.1, 0.2\}$), immortal populations were found not to be stable even on arrays of $2000\times 2000$ sites. Mortal populations were stable on $250\times 250$ arrays in all cases, with the exception of $g=0.005, v=0.2$ which required a $750\times 750$ array.

In most of the simulations presented in this work, we focus on the limiting case of no cost of reproduction ($c=0$), due to the greater clarity of interpretation in that case: e.g., in immortal populations, the reproductive restraint that evolves cannot be attributed to an individual effort to conserve resources, since reproducing is free to the individual. Increasing $c$ does not change the qualitative result that restraint is favored in the long term. We discuss the effect of nonzero values of $c$ further in the section on mean-field analysis below.

     In ascendance studies,
we initialized the lattice randomly with each site having a 55\% chance of being empty, a 40\% chance of having resources only, and a 5\% chance of having a consumer, with $p$ and $q$ in the latter case randomly chosen from a uniform distribution between 0 and 1.

\begin{figure}[tbp]
\begin{center}
\includegraphics[width=.5\textwidth]{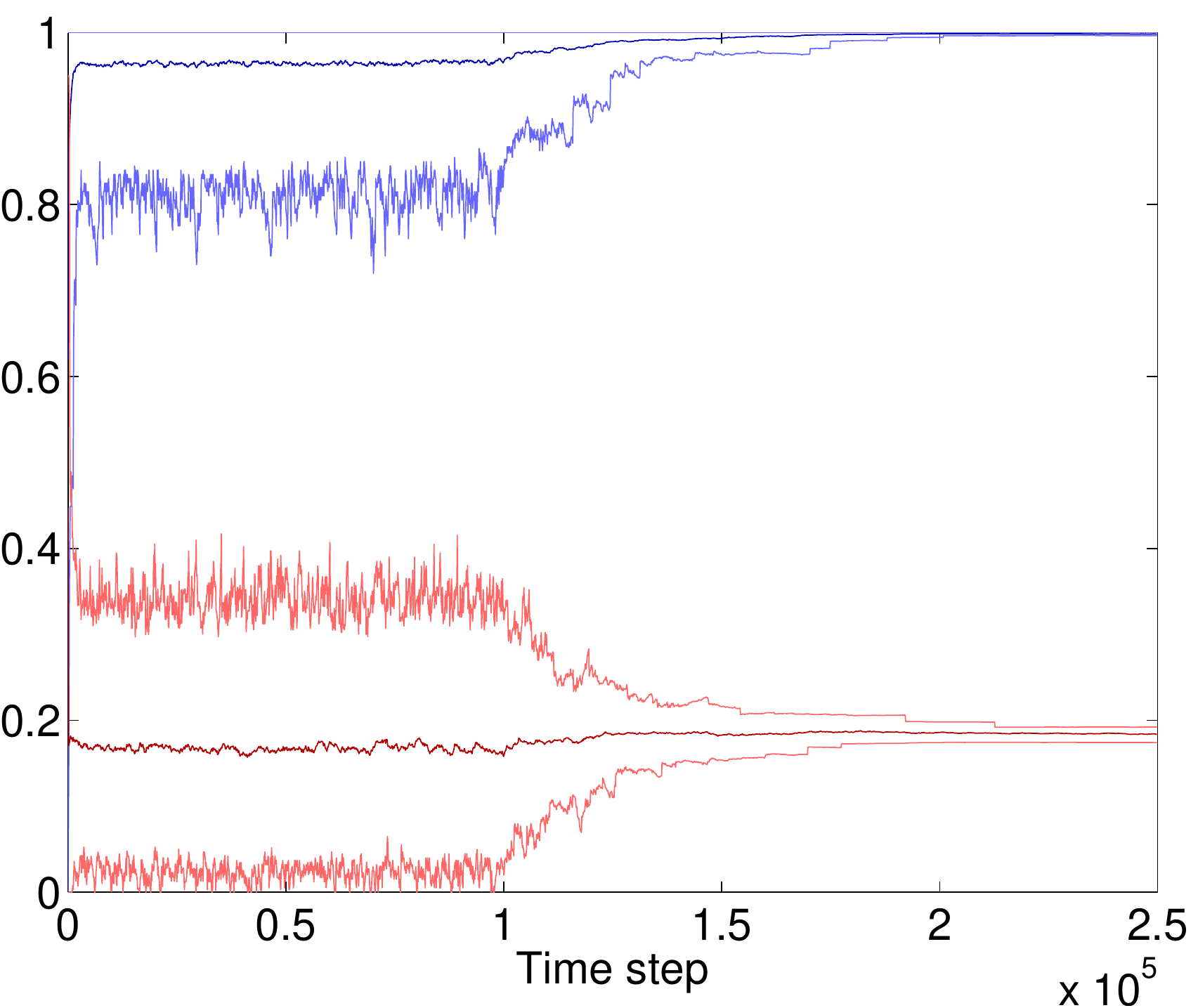}
\caption{Traces showing mean (dark) and maximum/minimum (light) values
  of $p$ (blue) and $q$ (red) over the full course of an ascendance
  study, with initial transient period ($t=0$--$100000$) and
  intermediate period during which $\epsilon_{p}$ and $\epsilon_{q}$
  are reduced ($t=100000$--$200000$).  Statistics reported in Fig.~3
  are based on the final $50000$ time steps of each simulation.  }
\label{reducing_epsilon}
\end{center}
\end{figure}

     Values of $p$ and $q$ are only meaningful within a finite range, $[0,1]$ and $[0,1-v]$ respectively. (The latter expression arises since starvation and intrinsic death are taken to be mutually exclusive possibilities.) Accordingly, mutants with values outside these ranges are set to the boundary values. To ensure that this operation does not cause artifacts in the steady-state value of $q$ or $p$, we performed simulations that progressively reduced the size of mutations $\epsilon_p$ and $\epsilon_q$. Following an initial 100,000 steps to achieve steady state, an additional 100,000 steps were performed during which $\epsilon_p$ and $\epsilon_q$ were halved every 10,000 steps, ending with a final 50,000 steps to obtain steady-state behavior (Fig.~\ref{reducing_epsilon}); the values reported in Fig.~\ref{fig:results} were based on this final period. The resulting minimum value of $q$, as well as the mean, was significantly above 0 in all cases in Fig.~\ref{fig:results}, supporting the finding that finite lifespan is consistently favored. 

In invasion studies, we initialized the simulation with a steady-state
configuration (following 200,000 time steps) of the ``invaded''
population. One consumer was then chosen at random and converted to an
invader, with $p$ unchanged, $q$ set to 0, and $\mu_q$ set to $0.1275$
or 0 according to whether the invader was mortal or immortal. If
invaders succeeded in taking over the entire consumer population, the
model was reset on the next time step to a steady-state configuration
of the invaded variant. If the invader's offspring became extinct, or
the steady state was reset following a successful takeover, a new
consumer was chosen at random and converted to an invader. The
simulation continued for 100,000 introduction attempts, or 500,000
attempts for immortals invading mortals (due to such an invasion's low
chance of success). In control studies in which both invader and
invaded were mortal, $q$ was left unchanged when an invader is
introduced.  Simulations for each $(g,v)$ pair were conducted on
arrays of the smallest size for which immortal consumer populations
could persist, as specified above. Table \ref{Table2} gives
probabilities of successful invasion for several values of $g$ and
$v$.

A modified version of invasion studies was used to evaluate the
long-term reproductive success of different types, as shown in
Fig.~\ref{timedep} (see also the discussion of ``time-dependent
fitness'').  The simulation was initialized with a steady-state
configuration for given values of parameters $g, v, c$; all consumers
had the corresponding equilibrium values of $p$ and $q$, and mutation
was turned off ($\mu_p=\mu_q=0$).  One consumer was chosen at random
and converted to an invader, and its offspring followed as in the
regular invasion studies above, with $q$ set to a new value for
invaders and $\mu_q$ still 0.  For the plots on the left side of
Fig.~\ref{timedep}, evaluating number of descendants as a function of
time, the initial invader was chosen from the entire consumer
population.  For the plots on the right-hand side of Fig.~1,
evaluating number of descendants as a function of generation number,
the initial invader was chosen from the set of newly-born consumers,
so that all its offspring were identified and assigned the correct
generation number.  The latter plots allow comparison of the number of
offspring after many generations for strains with different lifespans,
and are constructed as follows: each consumer in the invading
population has a generation counter, starting with 0 for the original
single invader, and with each offspring's counter being one higher
than that of its parent. The plots show the number of offspring of a
given generation number, no matter when in time they live.

\subsection*{Variations on the base model\label{si:var}}

In this section we briefly discuss how a number of qualitative changes to the base model affect its behavior. We consider: 
deterministic rather than stochastic limitations to lifespan (explicit programmed mortality, or rapid senescence);
increasing mortality with age (gradual senescence);
spatial mixing within a consumer population and migration of consumers within resource areas;
increased dispersal for reproduction of both consumers and resources;
deterministic resource use with continuous-valued resources, rather than binary resource state and stochastic consumption;
individual consumers' ability to adjust their ``rate of living'';
resources that regenerate spontaneously, rather than being spread only locally by existing resources;
consumer reproduction supplanting others already present;
resources that continue to regenerate even when exploited by consumers;
and sexual reproduction by consumers.

All these model variants support the generality of the conclusion that
finite lifespan is favored, when resources are limited and dispersion
is local.  For each variant, unless otherwise specified, we performed
10 independent simulations on a $250\times 250$ lattice with
$g=v=0.05$ and $c=0$.  Table \ref{Table3} shows collected results for
various cases.

\begin{sidewaystable}
\caption{{\bf Evolved trait values for selected model variants and
    conditions tested.}}
\begin{tabular}{|c|c|c||c|c|c|}
\hline
Variant & $g$ & $v$ & $p$ & $q$ & Notes \\
\hline\hline
\multirow{2}{*}{Reproduction cost} & 0.05 & 0.05 & 1 & $0.150\pm 0.002$ & $c=0.05$ \\\cline{2-6}
& 0.05 & 0.05 & $0.980\pm 0.007$ & $0.052\pm 0.001$ & $c=0.4$ \\\hline
Continuous-valued resources & 0.05 & 0.05 & 1 & $0.2770\pm 0.0009$ & -- \\\hline
Deterministic lifespan & 0.05 & 0.05 & 1 & $L=12.0\pm 0.1$ & -- \\\hline
\multirow{2}{*}{Senescence given by Gompertz equation} & 0.05 & 0.05 & 1 & $0.150\pm 0.003$ & $m_0=0.1$ (Gradual senescence) \\\cline{2-6}
& 0.05 & 0.05 & 1 & $-0.229\pm 0.001$ & $m_0=0.5$ (Negative senescence) \\\hline
Consumer rearrangement & 0.05 & 0.05 & 1 & $0.053\pm 0.001$ & -- \\\hline
Consumer migration & 0.05 & 0.05 & $0.984\pm 0.004$ & $0.0316\pm 0.0004$ & -- \\\hline
Rearrangement + migration & 0.05 & 0.05 & $0.96\pm 0.01$ & $0.053\pm 0.001$ & $7\times 7$ neighborhood \\\hline
Spontaneously generated resources & 0.005 & 0.05 & 1 & 0.95 & --\\\hline
Reproduction supplants existing consumers & 0.05 & 0.05 & 1 & $0.40\pm 0.05$ & -- \\\hline
\multirow{2}{*}{Increased consumer dispersal} & 0.05 & 0.05 & $0.985\pm 0.005$ & $0.24\pm 0.01$ & $R_R=1, R_C=1$ \\\cline{2-6}
& 0.05 & 0.05 & $0.934\pm 0.006$ & $0.16\pm 0.01$ & $R_R=1, R_C=2$ \\\hline 
\multirow{3}{*}{Increased resource dispersal} & 0.05 & 0.05 & $1$ & $0$ & $R_R=5, R_C=1$ \\\cline{2-6}
& 0.05 & 0.1 & $1$ & $0.11\pm 0.01$ & $R_R=5, R_C=1$ \\\cline{2-6}
& 0.05 & 0.2 & 1 & $0.402\pm 0.006$ & $R_R=5, R_C=1$ \\\hline
\multirow{4}{*}{Resources reproduce when exploited} & 0.05 & 0.23 & 1 & $0.25\pm 0.03$ & $g'=g$ \\\cline{2-6}
& 0.23 & 0.23 & 1 & $0.231\pm 0.002$ & $g'=0$ \\\cline{2-6}
& 0.05 & 0.05 & 1 & $0.0103\pm 0.0002$ & $g'=g/3$ \\\cline{2-6}
& 0.05 & 0.05 & 1 & 0 & $g'=g/2$ \\\hline
\multirow{2}{*}{Sexual reproduction} & 0.1 & 0.2 & $0.715\pm 0.004$ & $0.087\pm 0.001$ & Other parent chosen from $7\times 7$ neighborhood \\\cline{2-6}
& 0.1 & 0.2 & 1 & 0 & Other parent chosen from entire population \\\hline
Consumers adjust rate of living & 0.1 & 0.2 & 1 & $0.342\pm 0.001$ & $D=0.05, k=T=0.5$ \\
\hline
\end{tabular}
\begin{flushleft}  When resources are limited and dispersion is local, $q$ consistently evolves to a value significantly greater than 0. For comparison, in the base model when $g=v=0.05$ and $c=0$, $p$ evolves to 1 and $q$ to $0.1847\pm 0.0005$. Uncertainty measurements refer to standard error of the mean, based on 10 independent experiments. See text for details of variants.
\end{flushleft}
\label{Table3}
\end{sidewaystable}

\paragraph{Deterministic lifespan.}
In this variant, the genotype specifies intrinsic lifespan directly as a fixed length of time: $L$ time steps after a consumer is born, it dies, regardless of remaining resources.  Including some variability in the lifespan, by choosing it from some distribution with mean at $L$ and a moderate variance, does not change the results.

\paragraph{Increasing mortality with age.}
In this variant, a consumer's probability of death at each time step is given by the Gompertz equation \cite{Reznick04} $m(t)=m_0e^{qt}$, where $m_0$ is a constant (chosen here to be 0.1), $t$ is the number of time steps since the consumer's birth, and $q$ is a heritable value as in the base model.  In this case $q$ is not restricted by the simulation to be nonnegative; still, it evolves to a limited value.

Note that for other conditions, this model variant can predict the evolution of different senescence patterns. For instance, if intrinsic mortality is set to be initially much higher than the equilibrium found in the base model (e.g., $m_0=0.5$), then negative senescence can evolve.

\paragraph{Consumer migration.}
In this variant, we allow consumer mobility. For this purpose, we split each time step into three successive stages: first, reproduction of both resources and consumers; second, resource depletion and consumer death through both starvation and intrinsic mortality; and third, mobility, in which each consumer (asynchronously, in random order) is able to move.

We tested three types of mobility: (a) a consumer trades places with a
randomly chosen neighboring consumer (if any exist); (b) a consumer
moves to a neighboring, unoccupied resource site if such exists; (c) a
consumer exchanges places with a nearby randomly chosen non-empty site
of either type. In all three cases self-limited lifespan is favored.
Increasing the mobility distance to a limited extent (e.g., choosing
the target site in version (c) from a $7\times 7$ square centered at
the consumer's original position) does not change the qualitative
result.  Allowing consumers to move to a nearby resource site when
they exhaust the resources in their own site, in order to avoid the
starvation during that time step that would otherwise occur, likewise
does not change the qualitative result.

\paragraph{Increased consumer or resource dispersal.}
In this variant, consumers or resources are not limited to reproducing into only the four neighboring sites. For ease of implementation, a modified version of the model was used for these studies: Instead of the base model's synchronous update where each site was simultaneously updated based on its value and those of its neighbors, an asynchronous update was used in which sites were successively chosen in random order for updates of the following form: empty sites remained empty; resource sites remained resources, and had a probability $g$ of also reproducing by converting one empty site to a resource site; consumer sites had a probability $v$ of becoming an empty site and $q$ of becoming a resource site ($v+q\leq 1$), and separately a probability $p$ of reproducing by converting one resource site to a consumer site. For such resource or consumer reproduction, the offspring site was chosen randomly from all sites of the appropriate type such that both the row and column indices differed by at most $R_R$ or $R_C$ (for resources and consumers, respectively) from those of the parent. For $R_R=R_C=1$, this model produces the same qualitative results as the base model. (Note that these dispersal ranges are both effectively greater than those in the base model: reproduction is possible there only in the 4-neighborhood, here in the 8-neighborhood.)

     Increasing the consumer dispersal range $R_C$ can disrupt the local neighborhood relationships that make it possible for selection to favor restraint. Large enough $R_C$ makes the system effectively well-mixed. In such cases consumption increases through unchecked selection for faster reproduction and longer lifespan, until the consumer population exhausts all available resources and goes extinct. The dispersal range above which this occurs is a function of the ecological parameters $g, v, c$ as well as the size of the simulation space. This is because $g, v, c, R_C$ affect the length scale of the characteristic population structure; if that length scale is too large compared to the space, the spatial nature of the model breaks down. Thus a given value of $R_C$ can result in extinction on lattices of a given size but allow evolution of restraint and a sustainable consumer population on larger lattices. For instance, with $g=v=0.05$ and $c=0$, a consumer population with $R_C=1$ consistently goes extinct on a $50\times 50$ lattice (in $5800\pm 6300$ time steps), but survives (i.e., persists for at least $250,000$ time steps, evolving intrinsic mortality) on a $100\times 100$ lattice; $R_C=2$ consistently gives extinction on a $200\times 200$ lattice (in $60000\pm 37000$ time steps) but survival on a $250\times 250$ lattice; $R_C=3$ results in consistent extinction on a $250\times 250$ lattice (in $5400\pm 4900$ time steps), survival in 3 of 10 trials on a $500\times 500$ lattice (with extinction in $100000\pm 80000$ time steps in the other trials), and consistent survival on a $750\times 750$ lattice.

     Increasing the resource dispersal range $R_R$ increases the extent to which consumers generally have resources available, which affects the extent to which intrinsic mortality is favored. For instance, with $g=v=0.05$, increasing $R_R$ to 5 results in enough resource availability that intrinsic mortality is not favored; increasing the consumption rate $v$ to 0.1 reduces resource availability and returns the model to the regime where intrinsic mortality is favored, and increasing $v$ to 0.2 results in the evolution of still higher intrinsic mortality rates.

\paragraph{Continuous-valued resources and deterministic consumption.}
In this variant, we treat resources as continuous-valued and consumption as occurring with a fixed rate. All sites are characterized by the quantity of resources they contain, with a value from 0 to 1. Consumers deplete resources by an amount $v$ per time step, and $c$ per reproduction. When resources in a site reach 0, the consumer there dies. If the consumer dies prematurely due to intrinsic mortality, residual resources remain for future exploitation. When an empty site is converted to one containing resources, the resource value there is set to 1. Partially depleted resources are left unreplenished. Replenishing partially depleted resources slowly or quickly over time (to a maximum of 1) does not change the qualitative result.

\paragraph{Consumers can adjust rate of living in response to resource shortages.}
To consider the ability exhibited by many organisms to adjust rate of living in response to environmental conditions (e.g., dauer formation in \emph{C.\ elegans}, dietary restriction \cite{LongoFinch03,Kirkwood05}), we explored a variant based on the continuous-valued resource variant described above. In this variant, if a consumer is in a site with resource value below a threshold $T$, it adjusts its consumption $v$, reproduction $p$, and intrinsic mortality $q$ all by a multiplicative constant $k$. Additionally, depleted resources are gradually renewed, by an amount $D$ per time step, in both resource-only sites and those occupied by consumers.

\paragraph{Spontaneously generated resources.}
In this variant, we consider spontaneous appearance of resources at any location in space. This model corresponds, e.g., to certain plant-herbivore systems, where a plant such as grass can be cropped down to its roots and regrow. Empty sites in this variant become resource sites with probability $g$ at each time step independent of whether other resources are located nearby. Resources are generated much more readily than in the base model, and so a much lower value of $g$ produces an overall level of resource production more comparable to that in the base model. For high values of $g$, the consumer population is not limited by resource availability and lifespan control does not evolve.

\paragraph{Reproduction supplants existing consumers.}
In the base model, existing consumers prevent reproduction by neighbors into the sites they themselves occupy, in effect perfectly defending their territories.  The opposite extreme is for new challengers always to defeat older competitors. In this variant, we allow consumers to reproduce not only into sites with resources alone, but also those with consumers already present, replacing the previous consumer.

\paragraph{Resources reproduce even when exploited by consumers.}
In this variant, we explored the consequences of allowing resources to reproduce even when consumers are present---i.e., resource sites reproduce with probability $g$ in the absence of consumers and $g'$ in their presence (in the base model, $g'=0$).

     A nonzero $g'$ increases the overall level of resource availability. Hence this variant requires reducing $g$ and/or increasing $v$ to obtain results quantitatively comparable to those of the base model. For example, for $g=g'=0.05, v=0.23$, the steady-state value of $q$ is close to that of the base model for $g=v=0.05$ and for $g=v=0.23$. For a given $g$ and $v$, increasing $g'$ leads to lower steady-state values of $q$ (longer lifespan), just as increasing $g$ for fixed $v$ does in the base model (Fig.~3).

     It is possible in this variant, if the level of resource growth is high enough compared to $v$, for consumers to have little enough impact on the spatial distribution of resources that selection does not limit reproduction probability (for immortals) or lifespan (for mortals). For example, with $g=v=0.05$, reproduction and lifespan limits occur for $g'\leq g/3$ and not for $g'\geq g/2$.

\paragraph{Sexual reproduction by consumers.}
In this variant, when a consumer reproduces, a second consumer is randomly chosen to be the other parent of the offspring, with the values of $p$ and $q$ for the offspring being the average of those of the two parents plus mutation as in the base model. When the second parent is chosen from nearby (e.g., from a $7\times 7$ region centered on the offspring), finite lifespan is favored, as is reproductive restraint: $q$ evolves to a value significantly greater than 0, and $p$ to a value less than 1. When the second parent is drawn from the entire consumer population (i.e., a form of global dispersal), no restraint evolves: $q$ evolves to 0 and $p$ to 1. (In this latter condition, the changes in $q$ and $p$ in the population occur over a much longer time scale than in the base model, and population extinction follows the reduction of $\epsilon_p$ and $\epsilon_q$ described above. Delaying the start of that reduction allows $q$ and $p$ to reach those extreme values.)

\section*{Acknowledgments}
This work was supported by internal funding from the New England Complex Systems Institute (primarily 2007--2009), a DOD Breast Cancer Innovator Award (BC074986 to DEI), and the Wyss Institute for Biologically Inspired Engineering.

\end{document}